\documentclass[lettersize,journal]{IEEEtran}
\usepackage{cite}
\usepackage{amsmath,amssymb,amsfonts}
\usepackage{algorithmic}
\usepackage{graphicx}
\usepackage{textcomp}
\def\BibTeX{{\rm B\kern-.05em{\sc i\kern-.025em b}\kern-.08em
    T\kern-.1667em\lower.7ex\hbox{E}\kern-.125emX}}

\usepackage{xtab}
\usepackage{graphicx}
\usepackage{array} 
\usepackage{multirow} 
\usepackage[table,xcdraw]{xcolor} 
\usepackage{placeins} 
\usepackage{booktabs}    
\usepackage{ragged2e}
\usepackage{makecell}    
\usepackage{array}   
\usepackage[protrusion=true,expansion=true]{microtype}
\pdfoutput=1
\usepackage{tabularx}
\usepackage{url}
\usepackage[breaklinks=true]{hyperref}

\usepackage{placeins}  
\usepackage{float}     

\definecolor{Gray}{gray}{0.9}
\definecolor{darkyellow}{RGB}{249, 231, 159} 
\definecolor{salmon}{rgb}{1.0, 0.55, 0.41}
\definecolor{darkgreen}{RGB}{170, 230, 168}
\definecolor{lightgreen}{RGB}{209, 233, 201}

\definecolor{lightred}{RGB}{255,211,222}
\definecolor{darkred}{RGB}{252,150,167}
 \definecolor{LightBlue}{rgb}{0.8,0.89,1} 
\definecolor{DarkBlue}{rgb}{0.57,0.71,0.82} 
\definecolor{MagLight}{rgb}{1, 0.89, 0.8}
\usepackage{enumitem}

\begin{document}
\title{{\fontsize{23}{24}\selectfont  The Transition from Centralized Machine Learning to Federated Learning for Mental Health in Education: A Survey of Current Methods and Future Directions}}
\author{
Maryam Ebrahimi,~\IEEEmembership{Student~Member,~IEEE}, Rajeev Sahay,~\IEEEmembership{Member,~IEEE}, \\Seyyedali Hosseinalipour,~\IEEEmembership{Member,~IEEE}, and  Bita Akram,~\IEEEmembership{Member,~IEEE}\thanks{An abridged preliminary version of this work was accepted at the AI for Education - Tools, Opportunities, and Risks in the Generative AI Era (AI4EDU) Workshop held in conjunction with the 2025 AAAI Conference on Artificial Intelligence. Due to the non-archival nature of this workshop (with the paper not published in the conference proceedings), this preliminary version does not constitute a formal publication.
}\\
\thanks{
M. Ebrahimi is with Department of Mathematics, Sharif University of Technology, Tehran, Iran (e-mail: fatemeh.ebrahimi@sharif.ir). R. Sahay is with Electrical \& Computer Engineering Department, University of California San Diego, San Diego, USA (e-mail: r2sahay@ucsd.edu). S. Hosseinalipour is with Electrical Engineering Department, University at Buffalo--SUNY, Buffalo, USA (e-mail: alipour@buffalo.edu). B. Akram is with Computer Science Department, North Carolina State University, Raleigh, USA (e-mail: bakram@ncsu.edu). 
\vspace{-5mm}}}

\maketitle

\begin{abstract}
Research has increasingly explored the application of artificial intelligence
(AI) and machine learning (ML) within the mental health domain to enhance both patient care and healthcare provider efficiency. Given that mental health challenges frequently emerge during early adolescence -- the critical years of high school and college --  investigating AI/ML-driven mental health solutions within the education domain is of paramount importance. Nevertheless, conventional
AI/ML techniques follow a centralized model training architecture, which poses privacy risks due to the need for
transferring students' sensitive data from institutions, universities, and clinics to central servers. Federated learning (FL) has emerged as a  solution to address these risks by enabling distributed model training while maintaining data privacy. Despite its potential, research on applying FL to analyze students' mental health remains limited.
In this paper, we aim to address this limitation by proposing a roadmap for integrating FL into mental health data analysis within educational settings. We begin by providing an overview of mental health issues among students and reviewing existing studies where ML has been applied to address these challenges. Next, we examine broader applications of FL in the mental health domain to emphasize the lack of focus on educational contexts. Finally, we propose promising research directions focused on using FL to address mental health issues in the education sector, which entails discussing the synergies between the proposed directions with broader human-centered domains. By categorizing the proposed research directions into short- and long-term strategies and highlighting the unique challenges at each stage, we aim to encourage the development of privacy-conscious AI/ML-driven mental health solutions.

\end{abstract}
\vspace{-1mm}


\vspace{-2mm}
\section{Introduction}
\label{sec:introduction}
\noindent In recent years, mental health has gained recognition as a critical factor in achieving success and maintaining overall well-being in all areas of life, with its importance being particularly evident in educational settings~\cite{brannlund2017mental}. The World Health Organization (WHO) defines mental health as ``a state of well-being in which every individual realizes their own potential, can cope with normal stresses, works productively, and contributes to their community" \cite{world2004promoting}. This definition underscores mental health’s foundational role, not only in supporting students’ academic achievements but also in fostering holistic personal growth and resilience. Beyond the educational sphere, mental well-being is equally essential in other human-centered domains, such as professional work environments~\cite{lu2022relationship} and healthcare~\cite{huang2024role}, highlighting its broader significance in fostering well-rounded, productive individuals throughout  society.

\vspace{-2mm}
\subsection{Footprints of Machine Learning in Mental Health Analysis}
The importance of mental health becomes even more pronounced when considering its far-reaching consequences. In particular, contemporary research has shown that poor mental health can be the root cause of numerous serious personal and interpersonal issues, examples of which are social isolation, substance use disorders, and an increased risk of suicide \cite{wang2017social,forray2021collision}. More specifically, in educational contexts, students often face numerous stressors (e.g., heavy workloads, looming deadlines, and social expectations), which can hinder their ability to learn, perform, and succeed. As a result, institutions and educators are often interested in identifying the root cause of these stressors, which can signal adjustment of policies and teaching styles. Consequently, the early detection and treatment of mental health concerns -- both within educational settings and other human-centered domains -- have emerged as critical priorities for the research community~\cite{levitt2007early, pedrelli2015college,costello2016early}.

Yet, the quest to understand mental health and its complexities presents unique obstacles. Unlike physical health, which can be evaluated through objective measures like blood tests or imaging, mental health assessments rely heavily on subjective experiences and self-reported feelings. For example, while a broken bone can be confirmed through an X-ray that yields clear, universally interpretable evidence, diagnosing conditions such as depression or anxiety hinges on interpreting diverse, personal narratives of emotions and behaviors. This inherent subjectivity makes measuring and assessing mental health both challenging and nuanced. To navigate these difficulties, researchers traditionally employ three key methodological approaches to diagnosing mental health conditions: \textit{qualitative}, \textit{quantitative}, and \textit{mixed} methods~\cite{badu2019integrative}.
 In particular, qualitative methods tap into individuals’ subjective experiences through interviews, observations, and focus groups, offering rich insights into the lived realities of mental health~\cite{brown2001qualitative}. In contrast, quantitative methods provide a more standardized approach by collecting structured data -- such as survey responses and clinical assessments -- to objectively measure variables like depression severity or anxiety levels~\cite{creswell2017research}. By combining these approaches, mixed methods integrate the depth and context of qualitative findings with the rigor and generalizability of quantitative analysis, ultimately producing a comprehensive understanding of mental health~\cite{palinkas2011mixed}.

While each of these methodological approaches contributes important insights, quantitative methods play a particularly crucial role in mental health research. They employ statistical tools to analyze trends and relationships in structured data, often focusing on associations between mental health status and external factors (e.g., lifestyle, socioeconomic variables)~\cite{loewen2019lifestyle,tanaka2002examination,hautekiet2022healthy,strohschein2005household,costello2001poverty,perna2010impact}, prevalence rates of mental health disorders~\cite{whitney2019us,auerbach2018world,de2012prevalence,trollor2007prevalence}, the effectiveness of various treatments~\cite{elkin1989national,anker2009using}, and longitudinal changes in mental health over time~\cite{taylor2005changes,collishaw2004time}.
To facilitate such analyses, researchers typically convert qualitative responses into numerical data. For instance, a questionnaire might ask participants to rate how frequently they experience certain feelings using four response options: `Never', `Sometimes', `Often', or `Always.' Assigning numerical values to these categories (e.g., 0 = `Never', 1 = `Sometimes', 2 = `Often', 3 = `Always') allows for the calculation of an overall score by summing or averaging these values across related items. Clinical cut-off points can then translate these scores into meaningful categories -- such as minimal, mild, moderate, or severe levels of symptoms. This transformation of personal experiences into structured data enables more nuanced and statistically rigorous analyses, ultimately advancing the field’s understanding of mental health and developing better-informed strategies for intervention and support.

Although transforming qualitative data into numeric formats enables more nuanced and rigorous statistical analyses, conventional quantitative methods -- such as descriptive and inferential statistics~\cite{sutanapong2015descriptive} -- still face limitations when confronted with the complexity and dimensionality of modern mental health datasets. For instance, they may struggle to fully capture nonlinear relationships or intricate interactions among numerous variables, which are often inherent in mental health data. As a consequence, researchers are increasingly turning to more advanced analytical techniques capable of uncovering these subtle patterns and providing a more comprehensive understanding of the complex factors shaping mental well-being.
To this end, recent advancements in artificial intelligence (AI) and machine learning (ML) have emerged as powerful tools, which are capable of offering novel computational methods for uncovering intricate patterns hidden within the data.  In the mental health domain, studies have shown that AI/ML can assist with (i) early diagnosis of mental health issues~\cite{chung2022mental,iyortsuun2023review,ahmed2020machine,akinci2012video,reece2017forecasting}, (ii) providing personalized treatments \cite{talati2023artificial}, (iii) paving the way for public health enhancements~\cite{glasgow2014our,glasgow2016your}, and (iv) guiding the clinical decision-making processes \cite{bone2017signal,thieme2023designing}. Similarly, in the education domain, research has demonstrated the success of AI/ML techniques in (i) predicting academic variables (e.g., GPA, dropout rate)~\cite{rastrollo2020analyzing,kotsiantis2004predicting}, (ii) providing personalized analysis and feedback based on individual data~\cite{essa2023personalized,taylor2024personalized}, (iii) enabling automated grading systems~\cite{ndukwe2019machine,chen2024design}, and (iv) supporting students through chatbots~\cite{clarizia2018chatbot}. 
 
 \begin{figure}[t]
\vspace{-.5mm}
    \centering
    \includegraphics[width=0.47\textwidth]{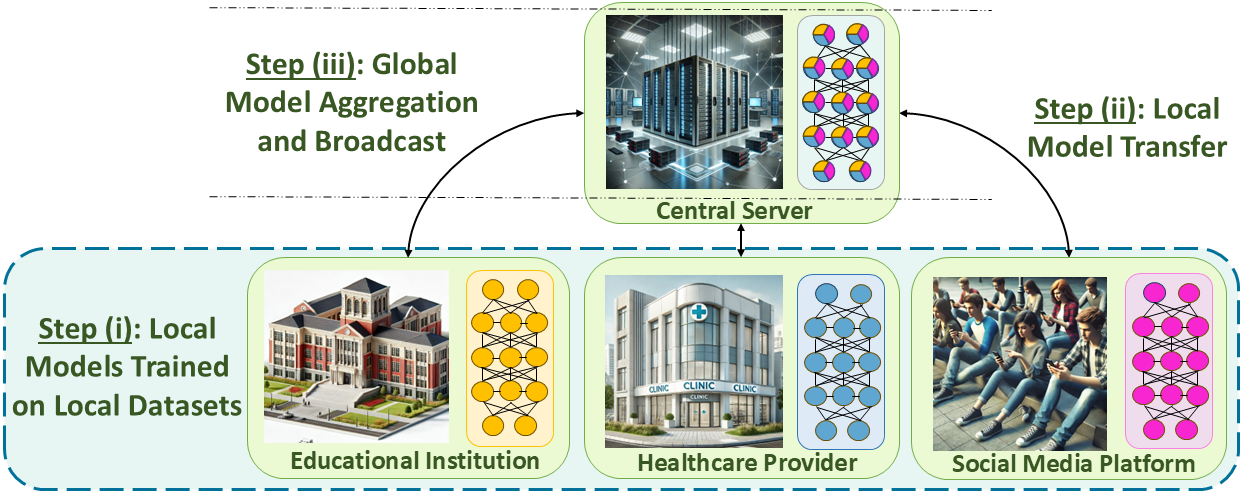}  
        \vspace{-3mm}
    \caption{A schematic of FL model training architecture.}
    \label{fig:diagram}
    \vspace{-5mm}
\end{figure}


 At the intersection of these domains, AI/ML shows promise in promoting students’ mental well-being by enabling the early detection of both mental health concerns and academic risks. For instance, ML techniques have been applied to predict students' mental health indicators, such as anxiety and depression, based on a combination of academic and lifestyle factors~\cite{nayan2022comparison,ahuja2019mental}. Similarly, these approaches have been used to forecast academic performance (e.g., GPA) using mental health metrics (e.g., stress levels, anxiety, and sentiment analysis), ultimately helping the identification of at-risk students and the development of personalized intervention plans~\cite{venkatachalam2023predicting}.

\vspace{-2mm}
\subsection{Migration from Centralized to Distributed ML Methods} \label{sec:intro_FL}
Despite the notable advancements of AI/ML in the mental health field, a common shortcoming persists across most existing studies: they typically rely on a \textit{centralized ML model training scheme}, in which models are trained on a central server that has direct access to aggregated data from multiple institutions. Such aggregation raises significant data privacy concerns, as sensitive mental health information could be compromised. In particular, the preservation of user privacy is a fundamental requirement in the mental health domain, making centralized approaches highly problematic.
Alternatively, focusing on data from a single institution -- such as one university or school -- can limit the performance and generalizability of AI/ML models due to exposure to a restricted dataset. For example, consider a university where international students represent only a small fraction of the population. Training a model solely on this university’s data may result in underrepresentation of these international students, potentially causing the model to perform poorly when predicting mental health indicators or academic outcomes for them. This highlights the importance of incorporating more diverse and \textit{geographically distributed (geo-distributed) data sources}, enabling richer data and improved model robustness. In the mental health context, such benefits are even more pronounced: data from academic institutions can be combined with data held at mental healthcare clinics, resulting in a more comprehensive perspective on mental well-being.

  Nevertheless, taking advantage of these geo-distributed datasets using traditional, centralized ML training methods is challenging. This is because current centralized ML model training approaches require that data be pooled from multiple, distributed data collection points and transferred to a central location (e.g., a cloud server) for model training. This data centralization not only exacerbates privacy risks but can also lead to violations of privacy regulations, some of which impose strict limitations on transferring even anonymized data~\cite{altman2021hybrid}. Additionally, recent research has shown that data anonymization does not always fully protect identities~\cite{ohm2009broken}, as re-identification attacks can sometimes recover sensitive information.
  

In response to these challenges, privacy-preserving distributed ML techniques have gained significant attention, among which the most prominent is \textit{federated learning (FL)}. FL (depicted in Fig.~\ref{fig:diagram}) aims to eliminate the transfer of raw data across networks. In particular, instead of aggregating data at a central server, FL involves local model training at each data collection point (e.g., universities, schools, mental health clinics) using their own datasets. Afterwards, only the trained local models, rather than the raw data, are shared with a central server. The server then aggregates the local models -- often through techniques like weighted averaging -- to form a global model, which it subsequently broadcasts back to the data collection points to initiate the next round of training~\cite{mcmahan2017communication}. This iterative process continues until the global model converges, thus preserving privacy by never exposing raw data.

\vspace{-2mm}
\subsection{Motivations and Contributions}

In the context of mental well-being, early research has begun to explore the application of FL for detecting mental health disorders~\cite{almadhor2023wrist}. However, given its emerging status, a comprehensive vision -- especially regarding FL’s applications at the intersection of mental health and education -- is still lacking. To address this gap, this work provides an overview of AI/ML methodologies for supporting students’ mental well-being and examines existing FL applications in mental health, culminating in the proposal of promising new research directions. Our contributions can be summarized as follows:



\begin{enumerate}[leftmargin=4.5mm]
    \item \textit{\textbf{Review of AI/ML Applications in Student Mental Health:}}
   We review the landscape of AI/ML applications in student mental health, uncovering significant efforts aimed at addressing issues such as stress, anxiety, depression, attention-deficit/hyperactivity disorder (ADHD), and substance use disorders (SUD). We then discuss that traditional centralized ML models raise privacy risks due to their reliance on aggregating sensitive student data at central servers. This review highlights the need for more privacy-conscious AI/ML approaches and their potential applications in student mental health analysis.

    \item \textit{\textbf{Exploration of FL in Mental Health Research:}}
 We then examine how FL has been applied in various mental health contexts, particularly for detecting stress, anxiety, and depression. We then discuss that despite promising results, its application in educational settings remains limited, leaving a critical gap in privacy-preserving mental health analytics in the education sector.

    \item \textit{\textbf{Survey of Relevant Mental Health Datasets:}}
    Recognizing the data-driven nature of FL, we survey relevant mental health datasets, categorizing them into three domains: educational, human-centered, and psychiatric.
     This dataset overview provides crucial insights into the types of data, ML tasks, accessibility, and distribution structures that can enable implementation of various FL methods in student mental health analysis.

    \item \textit{\textbf{Proposed Research Directions:}}
    We propose a set of promising research directions designed to push the boundaries of FL-driven mental health analysis for mental health analysis in the education domain. We further broaden the scope by examining how advancements in FL-driven student mental health research can benefit broader human-centered domains such as healthcare.
        In the short-term, we study applying FL across decentralized mental health datasets and integrating data with heterogeneous features. More importantly, looking further ahead, we proposed innovative approaches such as:
        \begin{itemize}[leftmargin=-.5mm]
            \item \textit{\textbf{Vertical FL (VFL):}} Integrating partial data from educational institutions, healthcare clinics, and online platforms.
            \item \textit{\textbf{Complementary Server-Side Learning:}} Combining centralized and decentralized datasets to create richer models.
            \item \textit{\textbf{Personalized FL (PFL):}} Customizing models to address data heterogeneity across different student populations.
            \item \textit{\textbf{Multi-Task FL (MTFL):}} Simultaneously predicting diverse mental health outcomes.
            \item \textit{\textbf{Large Language Models (LLMs):}} Deploying LLMs for mental health counseling within FL frameworks.
            \item \textit{\textbf{Explainable AI (XAI):}} Enhancing model interpretability to guide policy-making in educational institutions.
            \item \textit{\textbf{Multi-Modal FL (MMFL):}} Combining diverse data types such as text, images, and audio within FL models.
            \item \textit{\textbf{Federated Unlearning:}} Allowing specific data to be removed while retaining model utility.
            \item \textit{\textbf{Security and Privacy Enhancements:}} Exploring differential privacy, and cryptographic/security mechanisms.
            \item \textit{\textbf{Alternative FL Architectures:}} Investigating decentralized, semi-decentralized, and hierarchical FL frameworks to enhance scalability and model robustness.
            \item \textit{\textbf{FL under Data/Concept Drift:}} Adapting models to evolving mental health data through dynamic and online FL.
        \end{itemize}
\end{enumerate}

 Through this roadmap, we aim to lay a solid foundation for future exploration of FL-driven mental health analysis, ensuring both privacy protection and model transparency.

 \section{Major Mental Health Issues in Education (Anxiety, Stress, Depression, ADHD, and SUD)}
\label{sec:disorders}
\noindent  In this section, we provide an overview of major mental health issues in the education domain. In the later sections, we will demonstrate how these issues are targeted and addressed.

In the field of education, various mental health issues and their effects on student behavior and performance have been investigated. One significant mental issue among students is \textit{anxiety}, which involves feelings of nervousness and excessive worry, particularly when facing challenging tasks~\cite{american2013diagnostic}. This persistent worry can hinder students’ abilities to concentrate on academic tasks (e.g., lectures, assignments, and exams), while also disrupting their memory and information processing. These effects can significantly impair students' academic skills and undermine their overall performance~\cite{seipp1991anxiety}. 

Closely related to anxiety, \textit{stress} is another major factor influencing students' mental health and academic performance~\cite{sohail2013stress}. Although stress itself is not classified as a mental health disorder, chronic or excessive stress can severely impact cognitive functions, including concentration, memory, and problem-solving abilities. Over time, persistent stress can lead to the development of anxiety and depression, further compromising students' performance~\cite{checkley1996neuroendocrinology,patriquin2017neurobiological}.

\textit{Depression} is another mental health issue among students, characterized by persistent feelings of sadness, hopelessness, and a lack of interest in daily activities~\cite{american2013diagnostic}. This condition reduces motivation, making it difficult for students to study, complete assignments, or participate in extracurricular activities. Additionally, depression can drain students’ energies, making even routine tasks feel overwhelming. Research has also highlighted the negative impact of depression on students’ academic performance, stemming from these factors~\cite{hysenbegasi2005impact}.

\textit{Attention-deficit/hyperactivity disorder (ADHD)}, which involves enduring patterns of inattention, impulsivity, and hyperactivity~\cite{american2013diagnostic}, also impacts students' performance by impairing cognitive functions~\cite{weyandt2013performance}. Although ADHD often begins in childhood, it can continue into adulthood, affecting students' ability to maintain focus and remain calm in the classroom. This condition not only hinders students' performance but can also disrupt the learning environment for their peers.

Finally, \textit{substance use disorder (SUD)} -- defined as the inability to control the use of substances (e.g., alcohol or drugs)~\cite{american2013diagnostic} -- has increasingly become an issue among students. SUD weakens students' academic performance by impairing cognitive functions such as problem-solving ~\cite{mekonen2017substance}. Moreover, SUD can sometimes result from other mental health issues, such as depression and anxiety, compounding the negative effects on students' well-being and academic performance. 

Given the substantial impact of these mental health conditions on students' academic success, early detection and intervention are crucial to fostering a supportive educational environment and enabling students to reach their full potential. In the following section, we explore how ML techniques can be applied to effectively address these pressing issues.

\vspace{-1mm}
\section{Applications of ML  for Mental Health in Education} \label{sec:MLapp}
\noindent The integration of ML into mental health research has created new opportunities for accurately predicting and understanding mental health issues among students. In this section, we categorize current studies based on their application to specific mental health problems discussed in the previous section (Sec. \ref{sec:disorders}) and examine a selection of studies within each category. These studies have employed a wide range of algorithms, from traditional statistical models (e.g., logistic regression) to advanced neural networks, to diagnose mental health disorders and analyze factors influencing their development.

To provide a consolidated view of the studies discussed, we have summarized their key characteristics in Table~\ref{tab:MLpapers}. This table highlights each study's objectives, data domains, data modalities, and the ML methods used. By presenting this information collectively, Table~\ref{tab:MLpapers} illustrates the diversity of methodologies and applications of ML in diagnosing and understanding mental health issues among students.

\vspace{-2mm}
\subsection{ML for Stress Detection}\label{sec:MLstress}

A significant number of studies have explored the use of ML techniques to detect stress in students and identify its key predictors. In educational settings, it has been shown that students often experience heightened stress levels during specific periods, such as before and during exam seasons~\cite{vsimic2012exam,shrivastava2018assessment,parsons2008there}. Further, to analyze the impact of exams on students' stress levels, Ahuja and Banga~\cite{ahuja2019mental} utilized data collected from students' responses to the Perceived Stress Scale (PSS) questionnaire and applied several ML algorithms, including linear regression, naïve Bayes, random forest, and support vector machines, to classify stress levels into three categories: normal, stressed, and highly stressed. Additionally, they used these models to predict students' stress levels during internet usage, exploring its impact on students' mental health. A same trend of research has been carried in~\cite{pv2022classification,das2023stress, mayuri2020predicting}.

While traditional psychological tests such as the PSS are effective at capturing symptoms of stress, they primarily focus on the frequency and intensity of stress-related feelings. A more nuanced approach that incorporates a broader range of variables can enhance the accuracy and predictive power of the ML models. Motivated by this, Rois et al.~\cite{rois2021prevalence} integrated data on students’ academic performance, lifestyle habits, and physical health in their analysis. Using decision trees, random forests, support vector machines, and logistic regression, they achieved high prediction accuracy and identified significant predictors of stress, such as pulse rate, blood pressure, sleep patterns, smoking habits, and academic background. A similar methodology is pursued in~\cite{Sano2018-ed,aalbers2023smartphone,acikmese2019prediction}.





\vspace{-2mm}
\subsection{ML for Anxiety Diagnosis}

The application of ML to diagnose anxiety among students has garnered significant attention, driven by the growing prevalence of anxiety and the recent impact of the COVID-19 pandemic on students' mental health. Studies have shown that the pandemic significantly exacerbated students' anxiety levels, prompting the adoption of ML techniques to more accurately assess and predict these levels~\cite{liyanage2021prevalence,zhang2021anxiety,busetta2021effects,chang2021prevalence}. For instance, Wang et al.~\cite{wang2020chinese} employed the extreme gradient boosting (XGBoost) algorithm to evaluate anxiety among undergraduate students during the pandemic. By analyzing demographic data and responses to the Self-Rating Anxiety Scale (SAS) at two critical points -- before and after the transition to online learning -- the study revealed a marked increase in anxiety levels, highlighting the adverse effects of quarantine and remote learning on students' mental well-being.

Beyond standard psychological tests like the SAS, researchers have sought to gain more detailed insights into students' anxiety levels by developing specialized questionnaires tailored to their specific objectives. For example, Bhatnagar et al.~\cite{bhatnagar2023detection} created a questionnaire designed to identify both the causes and effects of anxiety, such as financial and academic stressors, as well as reduced joy and motivation. Using this dataset, the authors applied a variety of ML algorithms -- including naïve Bayes, decision trees, random forest, and support vector machines -- to classify students' anxiety into mild, moderate, and severe categories. 

As ML techniques have evolved, more sophisticated models have been employed to improve prediction accuracy and better address the complexities of anxiety detection~\cite{wang2019anxietydecoder,vulpe2021neural}. For instance, Devi et al.~\cite{devi2016adaptive} introduced an adaptive neuro-fuzzy inference system (ANFIS) to predict students' anxiety levels based on personality traits such as neuroticism and extroversion. These mental health indicators are inherently subjective, making it challenging to model their uncertainties and complex relationships using conventional approaches. By combining the adaptive learning capabilities of neural networks with the interpretability of fuzzy logic, ANFIS effectively captured these complexities, offering a robust and comprehensive framework for psychological research and anxiety prediction.


\vspace{-2mm}
\subsection{ML for Depression Diagnosis}
ML techniques have been widely applied to predict students' depression levels, supporting prevention and intervention strategies. For example, Zhai et al.\cite{zhai2024machine} utilized data from the Healthy Minds Study~\cite{hms} to develop predictive models using XGBoost, random forest, decision trees, and logistic regression. Their work aimed to enhance mental health services by identifying college students at higher risk of depression. 

Similar to the case of anxiety, the COVID-19 pandemic has significantly contributed to the rise of depression among students~\cite{quesada2024depression,zhang2021anxiety,azmi2022impact}. During the first wave of the pandemic, Nayan et al.~\cite{nayan2022comparison} employed various ML models to investigate depression in university students. Using an online questionnaire, they collected data on sociodemographic, educational, and socioeconomic factors, along with depression assessments, to evaluate the performance of models such as logistic regression, random forest, support vector machines, linear discriminant analysis, K-nearest neighbors, and naïve Bayes. They highlighted the importance of these factors in predicting depression, particularly during periods of heightened stress and uncertainty.

In addition to environmental and demographic influences, familial factors have been identified as critical predictors of depression in students~\cite{lopez1986family,zhao2018effects}. For instance, Gil et al.~\cite{gil2022machine} investigated the role of familial and personal factors in predicting depression among college students. By employing logistic regression, support vector machines, and random forest, their study highlighted the significant impact of family dynamics -- such as family cohesion, parental health conditions, and attachment styles -- alongside personal traits in shaping students' mental well-being.


\vspace{-2mm}
\subsection{ML for ADHD Diagnosis}
ADHD among students has been extensively studied, particularly in terms of its symptoms, prevalence, and effects~\cite{weyandt2006adhd,green2012we}. Despite this extensive literature, the application of ML models in the study of ADHD remains relatively limited. Nevertheless, some studies have applied ML techniques to predict ADHD in students -- particularly children -- and to explore its impacts and the factors influencing it. For example, Ren et al.~\cite{ren2021predicting} conducted a longitudinal study to investigate the persistence of ADHD and its influence on academic performance. By tracking boys diagnosed with ADHD from childhood into adulthood, the study employed stepwise multiple logistic regression on survey data and cognitive test results to identify key predictors, including IQ, family environment, and ADHD severity. These factors were found to play a critical role in shaping both clinical outcomes and academic performance over time.


\vspace{-2mm}
\subsection{ML for SUD Diagnosis}\label{sec:MLSUD}
To address substance use among students, ML techniques have been applied to identify high-risk behaviors among this population. For example, Marcon et al.~\cite{marcon2021patterns} conducted a study using elastic net, random forest, and neural networks to identify patterns of high-risk drinking and alcohol use disorders among medical students. By analyzing sociodemographic, personal, academic, and mental health data, the study developed predictive models for high-risk drinking and identified key associated factors within this population.
In particular, family factors, such as parental socioeconomic status and parental substance use, have been identified as significant predictors of substance use in students~\cite{sonmez2016substance,li2002parental}. To further analyze these predictors, Vázquez et al.~\cite{vazquez2020innovative} applied ML methods -- elastic net, K-nearest neighbors, random forest, and neural networks -- to data from a sample of southern American students. The dataset captured both individual factors (e.g., gender, grade, self-esteem) and socio-ecological factors (e.g., neighborhood quality, peer influences, parenting). Their findings emphasized peer and parental influences as key predictors, offering valuable insights for early identification and prevention strategies.

\vspace{-1mm}
\subsection{ML for Academic Performance Prediction Based on Mental Health State}\label{sec:MLACAD}
Building on the ability of ML to diagnose mental health issues among students, researchers have increasingly focused on understanding how these issues impact students' academic performance. At the same time, social media posts, containing rich personal and behavioral data, have become a valuable tool for assessing mental health status~\cite{garg2023mental,ganie2023social}. As a result, many studies have leveraged students' social media data to predict academic outcomes. For example, Venkatachalam and Sivanraju~\cite{venkatachalam2023predicting} extracted attributes related to students' mental health status and mood changes from their social media posts. They then proposed a deep neural network model to predict students' academic performance based on these attributes, combined with their academic and demographic data.
Similarly, to explore the relationship between students' mental health conditions and academic performance, Mukta et al.~\cite{mukta2022predicting} used neural networks to predict psychological attributes such as depression, self-efficacy, and personality traits from Facebook posts. These attributes were then used to predict students' GPA.



\vspace{-1mm}
\subsection{ML Limitations In Mental Health Study}

As we discussed in Sec.~\ref{sec:intro_FL}, privacy and ethical considerations remain a cornerstone of mental health research, especially when applying ML techniques. In particular, conventional centralized ML models which are discussed in the above subsections pose significant privacy risks due to their reliance on aggregating sensitive mental health data from diverse institutions to a central server. This approach raises concerns about data breaches and unauthorized access, making it challenging to ensure users' confidentiality and privacy.

FL emerges as a promising alternative by enabling decentralized ML model training while keeping sensitive data local to the institutions. In FL, raw data never leaves the local devices; instead, only model updates are shared with a central server for aggregation, minimizing privacy risks. Additionally, adhering to privacy and ethical research practices -- such as obtaining informed consent, anonymizing data, and securing approval from institutional review boards -- is expected to be easier in FL implementations as compared to centralized ML techniques. The adoption of FL, which is discussed in the following, thus represents a shift toward ethically sound AI/ML-driven mental health solutions that prioritize data privacy and compliance with evolving regulatory standards.


\begin{table*}[t]
    \centering
     \caption{Summary of studies applying ML techniques to students' mental health research.}
    
    \renewcommand{\arraystretch}{1.5}  
    \setlength{\tabcolsep}{8pt}        
    \begin{tabular}{>{\centering\arraybackslash}m{1cm}
                    >{\centering\arraybackslash}m{3cm}
                    >{\centering\arraybackslash}m{11cm}}
    \toprule
    \textbf{Reference} & \textbf{Data Domain/Modality} & \textbf{Objective}  \\

     \midrule
    \cite{nayan2022comparison}
    & Survey responses \newline (Textual and Numerical)
    
    & Predicting anxiety and depression status among university students based on demographic, academic, and socioeconomic factors using logistic regression, random forest, support vector machine, linear discriminant analysis, K-nearest neighbors, and naïve bayes algorithms. \\

    \midrule
    \cite{ahuja2019mental}
    & Survey responses \newline (Textual and Numerical)
    & Analyzing college students' stress levels at different time points based on PSS test responses using linear regression, naïve Bayes, random forest, and support vector machine algorithms. \\

\midrule
\cite{venkatachalam2023predicting}
& Institutional records(academic and demographic attribute) and social media text data 
\newline (Textual and Numerical)
& Predicting students' academic performance by analyzing academic and demographic data along with attributes related to their mental health status and mood changes extracted from social media posts using a deep neural network model. \\

    \midrule
    \cite{rois2021prevalence}
    &Survey responses \newline (Textual and Numerical)
    &Assessing stress prevalence among university students and predicting stress levels based on academic, lifestyle, health-related factors, and anthropometric measurements using decision tree, random forest, support vector machine, and logistic regression algorithms.\\

    \midrule
    \cite{wang2020chinese}
    & Survey resoponses \newline (Textual and Numerical)
    &  Predicting undergraduate students' anxiety levels during COVID-19 based on demographics and Self-Rating Anxiety Scale (SAS) responses using the XGBoost method.\\

    \midrule
    \cite{bhatnagar2023detection}
    & Survey responses \newline (Textual and Numerical)
    & Predicting university students' anxiety level based on research-specific questionnaire responses capturing anxiety causes (e.g., financial situation, academic factors) and anxiety effects (e.g., lack of joy) using naïve Bayes, decision tree, random forest, and support vector machine algorithms. \\

    \midrule
    \cite{devi2016adaptive}
    & Survey responses \newline (Textual and Numerical)
    &Predicting students' anxiety levels based on personality traits (e.g., neuroticism and extroversion) using adaptive neuro-fuzzy inference system method.\\

  \midrule
    \cite{zhai2024machine}
    & Survey responses\newline (Textual and Numerical)
    & Predicting anxiety and depressive disorder risks among college students using XGBoost, random forest, decision tree, and logistic regression algorithms applied to the Healthy Minds Study dataset.\\


    \midrule
    \cite{gil2022machine}
    & Survey responses \newline (Textual and Numerical)
    & Predicting depression levels of college students based on family factors using logistic regression, support vector machines, and random forest methods.
 \\

    \midrule
\cite{ren2021predicting}
& Survey responses and cognitive test outcomes \newline (Textual and Numerical)
& Predicting ADHD symptom persistence and academic outcomes (educational attainment) in adulthood for boys diagnosed with ADHD, based on childhood factors such as IQ, family environment, and ADHD severity, using stepwise multiple logistic regression. \\

\midrule
    \cite{marcon2021patterns}
    & Survey responses \newline (Textual and Numerical)
    & Identifying high-risk drinking and alcohol use disorder among students based on sociodemographic, personal, academic, and mental health factors using elastic net, random forest, and neural network.\\

  \midrule
    \cite{vazquez2020innovative}
    & Survey responses \newline (Textual and Numerical)
& Identifying key predictors of lifetime substance use (alcohol, tobacco, marijuana, inhalants) among students based on individual and socio-ecological factors using elastic net, k-nearest neighbors, random forest, and neural network.  \\

    \midrule
    \cite{mukta2022predicting}
    & Survey responses and social media text data \newline (Textual and Numerical)
    &Predicting students' mental health and psychological attributes (e.g., depression, self-efficacy, personality traits) from their Facebook posts using neural network (BiLSTM), and subsequently predicting their academic performance (GPA) based on these psychological and mental health attributes using an ensemble classification model. \\

    \bottomrule
    \end{tabular}
   
    \justifying
    \label{tab:MLpapers}
       \vspace{-3mm}
\end{table*}

\begin{table*}[t]
    \centering
    \caption{Summary of studies implementing FL approaches in the mental health domain.}
    \renewcommand{\arraystretch}{1.5}  
    \setlength{\tabcolsep}{8pt}        
    \begin{tabular}{>{\centering\arraybackslash}p{1cm}
                    >{\centering\arraybackslash}p{3cm}
                    >{\centering\arraybackslash}p{11cm}}
    \toprule
    \textbf{Reference} & \textbf{Data Domain/Modality} & \textbf{Objective}  \\

   \midrule
    \cite{almadhor2023wrist}
    & Physiological signals \newline (Numerical)
    &Predicting individuals' stress levels using deep neural network within FL framework applied to the WESAD dataset~\cite{schmidt2018introducing}.  \\

   \midrule
    \cite{fauzi2022comparative}
    & Physiological signals \newline (Numerical)
    &Predicting individuals' stress levels using logistic regression in individual, centralized, and FL frameworks applied to the WESAD dataset.  \\

    \midrule
    \cite{su2024classify}
    & Physiological signals \newline (Numerical)
    &Predicting workers' stress levels from physiological signals during human-robot interactions using centralized ML (support vector machine, multilayer perceptron, random forest, naïve Bayes) and FL (support vector machine) approaches.  \\
    \midrule
    \cite{shin2023fedtherapist}
    & Speech and keyboard input \newline (Textual and Auditory)
    & Designing a mobile mental health monitoring system that leverages deep neural network within FL for analysis of user-generated linguistic data (keyboard and speech inputs) to detect depression, stress, anxiety, and mood.\\
    
 \midrule
    \cite{gupta2024federated}
    & Physiological signals \newline (Numerical)
    & Predicting individuals' anxiety levels based on physiological data collected during virtual reality therapy sessions using adaptive ML models, including support vector machine and neural network, within a FL framework. \\

  \midrule
\cite{bn2022privacy}
& \hspace{4mm}Speech \newline (Auditory)
& Detecting depression based on speech features using deep neural network within FL framework applied to the DAIC-WOZ dataset.\\

       \midrule
   \cite{pranto2021comprehensive}
   & Keyboard input \newline (Textual)
   & Monitoring individuals' daily depression levels using local device text input, with a recurrent neural network trained within a FL framework.
\\

    \midrule
    \cite{tabassum2023depression}
    & Survey responses and  sensor data \newline (Textual and Numerical)
    & Predicting depression severity among students using PHQ-9 questionnaire responses and smartphone sensor data with neural network in both FL and centralized ML frameworks.\\

    \midrule
    \cite{qirtas2022privacy}
    & Survey responses and sensor data \newline (Textual and Numerical)
& Predicting loneliness using logistic regression, support vector machine, random forest, and XGBoost in both FL and centralized ML frameworks, applied to the StudentLife dataset~\cite{sls}. \\


    \bottomrule
    \end{tabular}
    \label{tab:FLpapers}
    
    \justifying
   \vspace{-3mm}
\end{table*}

\vspace{-2mm}
\section{FL for Mental Health Analysis} 
\label{sec:FLmental}
\noindent 
In the previous section (Sec. \ref{sec:MLapp}), we discussed the recent advancements and applications of ML in the mental health domain and the challenges posed by traditional, centralized ML approaches. FL has emerged as a promising alternative to address these challenges. By allowing data to remain within individual institutions, FL combines the analytical power of ML with strict adherence to data confidentiality requirements, reducing the privacy risks associated with raw data transmissions and the need for extensive data anonymization practices.

Despite its potential, FL is still in its early stages of adoption in mental health research compared to centralized ML frameworks. This limited adoption is understandable given the relatively recent emergence of FL as a viable technology. The existing studies that have leveraged FL in this field primarily focus on addressing prevalent mental health challenges, such as stress, anxiety, and depression, demonstrating FL's potential while also highlighting the narrow scope of its current applications. This research gap is even more pronounced in the educational context, where only a limited number of studies have used FL to examine students' mental health.
To provide a clearer understanding of the current state of FL research in mental health and identify existing gaps, in the following sub-sections, we categorize related studies based on their specific applications. Also, in  Table \ref{tab:FLpapers}, we summarize these studies, offering a comprehensive overview of FL's progress in the mental health domain.  

Given the extremely limited literature on FL applications for mental health analysis in education, the following sub-sections first explore FL's applications for mental health issues in broader human-centered domains (Secs.~\ref{sec:11},~\ref{sec:22}, and~\ref{sec:33}) and then review its current implementations in the education domain (Sec.~\ref{sec:FLedu}). This review serves as a foundation for motivating future directions of FL in mental health research for education and broader human-centered domains, which we will present in the next section (Sec.~\ref{sec:futureFL}).

\vspace{-2mm}
\subsection{FL for Stress Detection} \label{sec:11}
Stress often manifests through physiological changes, such as variations in heart rate and blood pressure, making the monitoring of these attributes essential for studying stress~\cite{kim2018stress,gasperin2009effect}. To mitigate the risk of data leakage when predicting stress levels from physiological attributes, researchers have explored the use of FL. For example, \cite{almadhor2023wrist, fauzi2022comparative} applied FL on the WESAD dataset, which contains physiological data collected using wearable devices.  In \cite{fauzi2022comparative},  Fauzi et al. evaluated logistic regression models under three configurations: individual-level training on each participant’s data, centralized training on aggregated data from all participants, and federated/distributed training within an FL framework. In their FL framework, each participant acted as a client, processing their data locally rather than sending it to a central server. While the FL model did not achieve the same accuracy as the centralized methods, it provided significant privacy advantages over the centralized approach. To improve prediction accuracy, Almadhor et al. \cite{almadhor2023wrist} employed a deep neural network on the same dataset, demonstrating FL’s ability to balance performance and privacy in predicting stress. They split the WESAD dataset into two subsets, each representing a simulated client, to create a distributed environment suitable for applying FL.

FL has also been applied to detect stress in workplace environments. Su et al. \cite{su2024classify} collected physiological data from participants engaged in human-robot interaction tasks. Initially, they evaluated various ML algorithms -- including support vector machines, multilayer perceptrons, random forests, and naïve Bayes -- within a centralized setup, with support vector machines achieving the highest accuracy. They then implemented this model in FL, which resulted in a high-performance, privacy-preserving solution for detecting workplace stress. 

\vspace{-2mm}
\subsection{FL for Anxiety Diagnosis}\label{sec:22}
FL holds significant promise for anxiety prevention and intervention by enabling privacy-preserving assessments of individuals' anxiety levels, particularly in contexts where data is inherently distributed. Smartphones, for example, which generate a wide variety of data, have become valuable distributed data resources for mental health analysis in recent years~\cite{beames2024use,boonstra2018using}. However, aggregating data from multiple smartphones raises serious privacy concerns, which FL can effectively mitigate. For instance, Shin et al.~\cite{shin2023fedtherapist} proposed an FL-based system to monitor anxiety through linguistic data such as speech and keyboard inputs. Through a custom application, they collected data over a 10-day period and applied deep neural networks to analyze participants' anxiety levels alongside other mental health indicators.

Beyond monitoring, FL has also been explored as a tool for supporting therapeutic interventions. For example, Gupta and Ekström \cite{gupta2024federated} combined FL with adaptive ML models -- including support vector machines and recurrent neural networks -- to develop a privacy-preserving system for detecting anxiety during virtual reality (VR) therapy sessions. Their approach utilized physiological signals (e.g., heart rate, skin conductance, and eye-tracking metrics) collected from participants in VR scenarios to enable real-time anxiety detection.

\vspace{-3mm}
\subsection{FL for Depression Diagnosis} \label{sec:33}
To assess individuals' depression levels, researchers have utilized a diverse range of distributed data sources, including self-rated questionnaires, interviews, and social media posts. Interviews, in particular, serve as a valuable source of information, offering flexibility in examining individuals' depression status~\cite{freedland2002depression,guohou2020reveals}. Addressing privacy challenges associated with the pooling of distributed data, Suhas BN and Abdullah \cite{bn2022privacy} explored the use of FL for privacy-preserving speech analysis to assess depression. Using the DAIC-WOZ dataset, which contains audio interviews to evaluate participants' mental health, they partitioned the dataset across multiple clients to simulate a decentralized setup. They fine-tuned pre-trained deep neural networks within the FL framework and demonstrated that the FL models achieved comparable accuracy to centralized models, highlighting its practicality for such applications.

Expanding beyond traditional diagnostic tools (e.g., questionnaires and interviews), Pranto and Al Asad \cite{pranto2021comprehensive} explored the use of FL to analyze social media data for depression monitoring. They utilized smartphone keyboard input data to track depression on a daily basis, employing recurrent neural networks within an FL framework. Their study illustrated how FL could be deployed on distributed personal devices, enabling privacy-preserving monitoring of individuals' mental health.

\subsection{FL for Mental Health Analysis in Education}
\label{sec:FLedu}
The potential of FL to address students' mental health has been explored in only a few studies so far. 
In~\cite{tabassum2023depression}, Tabassum et al. leveraged FL for diagnosing depression by developing an Android application that collected students' smartphone data. By pairing this data with students' responses to the PHQ-9 questionnaire, they utilized a deep neural network integrated with FL to assess depression levels. In a related study, Qirtas et al.~\cite{qirtas2022privacy} investigated the use of FL to detect loneliness among students. Using the StudentLife dataset in combination with responses to the UCLA Loneliness Scale questionnaire, they applied various ML models --including logistic regression, support vector machines, random forests, and XGBoost -- within an FL framework to classify students’ loneliness status.

These studies have taken the first steps in unveiling FL’s capacity to deliver privacy-preserving mental health assessments in educational settings. Yet, they also emphasize the need for further exploration into how FL can be customized to meet students’ unique needs.


\section{Future Directions: FL for Mental Health Analysis From Education to Broader Human-Centered Domains} \label{sec:futureFL}
\noindent As discussed in  the previous section (Sec.~\ref{sec:FLmental}), the implementation of FL for mental health analysis, particularly in the education domain, is still in its early stages. To fully explore FL's potential in addressing mental health challenges in education and its synergies with broader human-centered domains, it is essential to understand the available datasets for training and evaluating such models. Therefore, in this section, we first present a selection of relevant mental health datasets that can be leveraged in FL frameworks (Sec.~\ref{sec:datasets}). We then propose a series of open research directions containing a short-term vision (Sec.~\ref{sec:short}), which can be construed as natural extensions of FL techniques to this area, and several long-term visions (Sec.~\ref{sec:VFL} to Sec.~\ref{sec:drift}), which entail exploration of novel areas of research in this domain.

\subsection{Overview of the Relevant Mental Health Datasets} \label{sec:datasets}
We provide a summary of the relevant datasets containing mental health data in the educational domain and other related human-centered domains in Tables \ref{tab:datasets} and~\ref{table:IV}. Color coating is used to enhance the readability where shades of blue in  Table~\ref{tab:datasets} capture the datasets in the \textit{education domain}, shades of green in  Table~\ref{table:IV} indicate datasets in broader \textit{human-centered domains}, and shades of red in Table~\ref{table:IV} highlight datasets involving \textit{psychiatric} data.
We note that datasets in the mental health domain are either \textit{inherently centralized} or \textit{inherently decentralized/distributed} (denoted by the \textit{Collection} column in the tables). Inherently decentralized/distributed datasets often involve data collection from multiple institutions, while inherently centralized datasets often contain the data collected and stored in a single institution.
Although decentralized datasets are sometimes stored centrally for researchers to train AI/ML models, they are representative of the types of studies that lead to decentralized data collection, centralization of which is often not permitted due to privacy concerns. In Sec.~\ref{sec:comp}, we will further unveil research directions where centralized datasets can be used in conjunction with decentralized datasets in FL. The tables further summarizes the \textit{features} in datasets, the \textit{labels} of datapoints, \textit{potential ML tasks} that each dataset can be used for, and the \textit{accessibility}.   


\subsection{\texorpdfstring{\underline{\textbf{Short-Term Vision:}}}{Short-Term Vision:} Conventional FL Applications} \label{sec:short}


Inspecting Tables \ref{tab:datasets} and~\ref{table:IV}, several of the datasets are inherently decentralized, such as the Healthy Mind Study, National College Health Assessment, Longitudinal Study of
Australian Children, Add Health, University Students Mental Health, and Australian Student Performance, which incorporate data from various institutions. Additionally, the StudentLife dataset aggregates data from individual students' smartphones, while the National Comorbidity Survey, Adolescent Brain Cognitive Development, and UK Biobank dataset has been collected from different research institutions, and the Nurse Stress Prediction Wearable Sensors dataset involves data collected from nurses' wearable devices. This distributed characteristic presents an opportunity to apply conventional FL methodologies \cite{zhang2021survey}, providing the first privacy-preserving distributed ML benchmarks in the area. In particular, FL can be studied for stress detection, and anxiety, depression, ADHD, and SUD diagnoses using the metrics developed in the studies summarized in Sec.~\ref{sec:MLstress} to Sec.~\ref{sec:MLSUD} for centralized ML methods and in Sec.~\ref{sec:11} to Sec.~\ref{sec:33} for FL applications in generic human-centered domains. Further, FL can be adapted to the methods summarized in Sec.~\ref{sec:MLACAD} for academic performance prediction.
An alternative short-term research direction involves the application of FL across heterogeneous datasets, effectively merging them, while accounting for the \textit{inherent non-uniformities} among datasets, including differences in feature spaces and sample sizes.



\begin{table*}[!htbp]
\centering
\caption{Summary of the relevant datasets in the education domain along with their features, labels, potential ML tasks, accessibility, and their
data collection procedure.}
\begin{tabular*}{\textwidth}{| > {\centering\arraybackslash}m{0.1cm}
  || >{\centering\arraybackslash}m{2cm}X
  | >{\centering\arraybackslash} m{3.5cm}X
  | >{\centering\arraybackslash}m{2.7 cm}X
  | >{\centering\arraybackslash}m{3.5cm}X
  | > {\centering\arraybackslash}m{1.6cm}
  | >{\centering\arraybackslash}m{1.6 cm}|}
  \hline

\multicolumn{1}{|c||}{\cellcolor{black!10}} & \cellcolor{darkyellow} \textbf{ Dataset Name} & \cellcolor{darkyellow}\textbf{Features} & \cellcolor{darkyellow}\textbf{ Labels} &\cellcolor{darkyellow} \textbf{ ML Tasks} & \cellcolor{darkyellow}\textbf{ Accessibility} & {\cellcolor{darkyellow}\textbf{Collection}} \\ \cline{2-7}
\hline
\hline

     \multirow{10}{*}{\parbox[c][15.5 cm][c]{0.1 cm}{ \centering \rotatebox{90}{\textbf{ Education}}}}   & \cellcolor{LightBlue}  Add Health \cite{ah} &  \cellcolor{LightBlue} Demographics, physical and mental health factors, social and environmental factors
, behavioral factors, academic factors
   & \cellcolor{LightBlue} Mental health status, academic performance    &  \cellcolor{LightBlue} Predicting students' mental health status or academic performance, clustering students based on similar mental or academical characteristics    & \cellcolor{LightBlue} Public-use datasets available/restricted-use data available by request  &  \cellcolor{LightBlue} Decntralized  \\ \cline{2-7}
   
       &
   \cellcolor{DarkBlue}Australian Student Performance \cite{asp}   & \cellcolor{DarkBlue}Demographics, academic factors, environmental factors, social factors, physical and mental health factors   & \cellcolor{DarkBlue}Mental health status, academic performance  & \cellcolor{DarkBlue}Predicting students' mental health status or academic performance   & \cellcolor{DarkBlue}Available  & \cellcolor{DarkBlue}Decentralized  \\ \cline{2-7}

    & \cellcolor{LightBlue}Longitudinal Study of Australian Children \cite{lsac}   & \cellcolor{LightBlue}Demographics, physical and mental health factors, academic factors, family 
environment, social and environmental factors, emotional development  & \cellcolor{LightBlue}Mental health status, academic performance, socioeconomic 
outcomes
(e.g., household income)   & \cellcolor{LightBlue}Predicting students' mental health status, academic performance, or socioeconomic status, clustering students based on similar mental or academical characteristics   & \cellcolor{LightBlue}Available by request  & \cellcolor{LightBlue}Decentralized  \\ \cline{2-7}

    & \cellcolor{DarkBlue} National College Health Assessment\cite{ncha}   & \cellcolor{DarkBlue}Demographics, physical and mental health indicators, behavioral factors, academic factors,
safety and violence, environmental factors & \cellcolor{DarkBlue}Mental health status, academic performance    & \cellcolor{DarkBlue}Predicting students' mental health status or academic performance, clustering students based on similar mental or academical characteristics   &\cellcolor{DarkBlue} Available by request &\cellcolor{DarkBlue} Decentralized    \\ \cline{2-7}

& \cellcolor{LightBlue}Student-Depression-Text \cite{sdt}& \cellcolor{LightBlue}Age, gender, students' posts in social media  & \cellcolor{LightBlue}Anxiety or depression indicator    & \cellcolor{LightBlue}Precticting students' mental health status based on their social media posts  & \cellcolor{LightBlue}Available &\cellcolor{LightBlue} Centralized    \\ \cline{2-7}

   & \cellcolor{DarkBlue}StudentLife Study \cite{sls}   & \cellcolor{DarkBlue}Smartphone interaction, study duration, activity level, academic factors,
communication frequency, location pattern, self-reported stress level & \cellcolor{DarkBlue} Mental health status, academic performance    & \cellcolor{DarkBlue} Predicting students' mental health status or academic performance, clustering students based on similar mental or academical characteristics    & \cellcolor{DarkBlue}Available  & \cellcolor{DarkBlue}Decentralized \\ \cline{2-7}

    &\cellcolor{LightBlue} Student Stress Factors: A Comprehensive Analysis \cite{ssf} & \cellcolor{LightBlue}Physical and mental health factors, environmental factors, academic factors, social factors  &\cellcolor{LightBlue} Stress level, academic performance    &\cellcolor{LightBlue} Predicting students' mental health status   & \cellcolor{LightBlue}Available &\cellcolor{LightBlue} Centralized    \\ \cline{2-7}

     & \cellcolor{DarkBlue} The Healthy Mind Study \cite{hms}  & \cellcolor{DarkBlue}Demographics, physical and mental health indicators, help-seeking, behavioral factors, environmental factors, academic factors, public policing, financial stress   & \cellcolor{DarkBlue} Mental health status, academic performance       & \cellcolor{DarkBlue}Predicting students' mental health status or academic performance, clustering students based on similar mental or academical characteristics     & \cellcolor{DarkBlue}Available by request  & \cellcolor{DarkBlue} Decentralized \\ \cline{2-7}

    & \cellcolor{LightBlue}University Students Mental Health \cite{usmh}& \cellcolor{LightBlue}Age, gender, university, department, academic year, scholarship, mental health indicators  &  \cellcolor{LightBlue}students' depression, stress, and anxiety level   & \cellcolor{LightBlue}Predicting students' mental health status    &\cellcolor{LightBlue} Available   & \cellcolor{LightBlue}Decentralized \\ \cline{2-7}



    &  \cellcolor{DarkBlue}What Makes a University Student Life "Ideal"? \cite{isl}  &  \cellcolor{DarkBlue}Demographics, university-related factors, stress level    &  \cellcolor{DarkBlue}stress level, satisfaction level    &  \cellcolor{DarkBlue}Predicting students' stress levels or satisfaction level   &  \cellcolor{DarkBlue}Available  & \cellcolor{DarkBlue}Centralized  \\ \cline{2-7}

     

  \hline 

\end{tabular*}

\label{tab:datasets}
   \vspace{-5mm}
\end{table*}

\subsection{\underline{\textbf{Long-Term Vision 1:}} Vertical FL (VFL)}\label{sec:VFL}

VFL is designed for FL scenarios where multiple data collectors  hold different features for the same individuals (i.e., each individual has a common label but its data is distributed across institutions)~\cite{liu2024vertical}. 
In the mental health domain within educational settings, features/attributes of data of each student can be distributed across various domains: academic records held by schools, psychiatric records by clinics, and online behavior by social media sites. Given the sensitivity of data, students may be reluctant to allow it to be transferred across the platforms. VFL naturally unlocks \textit{the moonshot goal of collective usage of this distributed data (clinical, educational and online), for mental health analysis.} 
Although this research direction is highly enticing, to our knowledge the vertical datasets, records of the same students across different domains (clinics and education institutions), are not yet publicly available. Nevertheless, potentials of VFL can be studied through collaborations across various domains (e.g., ML specialists, education institutions and clinics) to unlock the access to such vertical data of students.

Further, despite its potentials, applications of VFL in the mental health domain within educational settings requires consideration of various nuances. Variability in data collection methods (i.e., different feature extraction techniques) across organizations can lead to incompatibility among features, which ultimately hinders VFL's effectiveness. Also, institutions (e.g., schools and clinics) often have different feature space sizes, leading to heterogeneities that could bias models toward institutions with larger feature spaces. To combat these, standardization techniques for feature extraction, such as feature alignment and harmonization frameworks~\cite{yu2021fed2,whitney2020harmonization}, could be proposed to ensure consistency across datasets. Further, applying transfer learning \cite{tan2023transfer} or multi-task learning \cite{aoki2022heterogeneous} can help adjust for varying feature space sizes, ensuring the model performs consistently across all institutions. 






\subsection{\underline{\textbf{Long-Term Vision 2:}} FL with Complementary Learning} \label{sec:comp}

In FL, training an ML model relies on decentralized datasets. However, many centralized datasets exist in practice. 
Thus, one of the moonshot research directions in the mental health domain within educational settings is to unlock the potential of \textit{combining centralized and decentralized datasets} -- especially when these datasets overlap in their ML tasks and feature space. 
From a practical standpoint, this can be accomplished by extending FL with a \textit{complementary server-side learning} approach, where the central server can hold its own central dataset \cite{mai2022federated}. To obtain prototypes, centralized datasets in Table~\ref{tab:datasets} -- such as  Student Stress Factors -- could be hosted at a central server, while decentralized datasets from individual schools or universities are maintained at the institutions. In this setup, the server acts as a client, training the model on its own data after aggregating updates from other clients. This results in a richer model that captures a wider variety of information.

 The concept of FL with complemented server-side learning can be further applied to both educational and psychiatric datasets in Tables \ref{tab:datasets} and~\ref{table:IV}. In this framework, decentralized educational datasets from local schools can be combined with clinical data stored centrally by a local healthcare provider. For instance, a decentralized educational dataset such as StudentLife could be paired with clinical data from the DAIC-WOZ dataset. After aggregating updates from the local models containing insights into students' educational data, the server would then train the global model on the students' clinical data, capturing additional important features related to their mental health. Studying the performance of FL  with a complementary server-side learning in the mental heal domain in terms of accuracy and fairness and other metrics of interest is thus a promising avenue of exploration.

\subsection{\underline{\textbf{Long-Term Vision 3:}} Personalized and Multi-Task FL} \label{sec:PFL}
Training a unified global model in FL often leads to  bias, favoring majority populations, where the model may overlook the unique characteristics of \textit{minority groups} -- a major issue  in human-centered domain. Personalized FL (PFL) aims to address this issue through tailoring models to individual clients/groups considering their specific data distributions via the use of meta-learning approaches or the use of a personalization layer during the model training procedure. In PFL, each client ends up with a distinct unbiased local model that is \textit{fine-tuned} to its specific dataset \cite{tan2022towards,arivazhagan2019federated}.
The adaptation of PFL in the mental health domain within educational settings has largely been overlooked. In the educational context, regional differences in social norms, family expectations, and cultural expressions of mental health results in non-IID (non-Independent and Identically Distributed) mental health data across institutions. This variability can bias the model toward the majority population in conventional FL, violating its fairness and weakening its overall performance. By applying PFL in the mental health domain, this issue can be effectively addressed. Further, the study of fairness metrics, such as \textit{statistical parity difference, disparate impact, equalized odds, average odds difference, and calibration and balance for positive/negative class}~\cite{selbst2019fairness} after applying PFL on various datasets in this domain is an unexplored area of research.

A related concept is  multi-task FL (MTFL), which is unexplored in the mental health domain within educational settings. MTFL addresses cases where different clients possess distinct but related ML tasks. For example, institutions (e.g., schools) may have distinct objectives -- one school might aim to predict students' stress levels, while another may focus on predicting students' depression levels. MTFL facilitates this by allowing each institution to leverage insights from other institution that are relevant to their
specific objectives while maintaining specialized models tailored to their tasks~\cite{smith2017federated}. 
As the number of tasks and institutions increase, MTFL can be combined with clustering techniques to capture task similarities and help models leverage insights from institutions with related objectives. However, identifying these similarities -- especially in human-centered domains -- need to be carefully studied.

\begin{table*}[!t]
\centering

\caption{Summary of the relevant datasets in general human-centered and psychiatric domains along with their features, labels, potential ML tasks, accessibility, and their data collection procedure.}\label{table:IV}
\begin{tabular*}{\textwidth}{| > {\centering\arraybackslash}m{0.1cm}
  || >{\centering\arraybackslash}m{2cm}X
  | >{\centering\arraybackslash} m{3.5cm}X
  | >{\centering\arraybackslash}m{2.7 cm}X
  | >{\centering\arraybackslash}m{3.5cm}X
  | > {\centering\arraybackslash}m{1.6cm}
  | >{\centering\arraybackslash}m{1.6 cm}|}
  \hline

\multicolumn{1}{|c||}{\cellcolor{black!10}} & \cellcolor{darkyellow} \textbf{ Dataset Name} & \cellcolor{darkyellow}\textbf{Features} & \cellcolor{darkyellow}\textbf{ Labels} &\cellcolor{darkyellow} \textbf{ ML Tasks} & \cellcolor{darkyellow}\textbf{ Accessibility} & {\cellcolor{darkyellow}\textbf{Collection}} \\ \cline{2-7}
\hline
\hline

\multirow{9}{*}{\parbox[c][14 cm][c]{0.1cm}{\centering \rotatebox{90}{\textbf{General Human-Centered Domains}}}}

  &\cellcolor{lightgreen} AnnoMI  \cite{9746035}  & \cellcolor{lightgreen} Therapy sessions   & \cellcolor{lightgreen}Quality of counseling, therapist and client attributes, personalized advice, sentiment analysis  & \cellcolor{lightgreen}Predicting individuals' mental health status and sentiment analysis, predicting personalized advice  & \cellcolor{lightgreen}Available  &\cellcolor{lightgreen}Centralized  \\ 
  \cline{2-7}

   &\cellcolor{darkgreen}CLPsych 2015  \cite{clp}  & \cellcolor{darkgreen}Text data from Twitter   & \cellcolor{darkgreen}Mental health status, sentiment
analysis&\cellcolor{darkgreen} Predicting users’ mental health status and sentiment
analysis  &\cellcolor{darkgreen} Available by request  &\cellcolor{darkgreen}Centralized  \\ \cline{2-7}

    & \cellcolor{lightgreen} eRisk   \cite{erisk}  & \cellcolor{lightgreen} Text data across various social media &\cellcolor{lightgreen}  Mental health status, sentiment analysis   & \cellcolor{lightgreen} Predicting individuals' mental health status or sentiment analysis based on their social media posts
health condition categories    & \cellcolor{lightgreen} Available by request &\cellcolor{lightgreen} Decentralized  \\ \cline{2-7}

 &\cellcolor{darkgreen} Mental Health Support Feature Analysis \cite{mhs} & \cellcolor{darkgreen} Text data from Reddit-  indicators measure (Automated Readability Index, Coleman-Liau Index,...), TF-IDF analyses, text metrics &\cellcolor{darkgreen} Mental health status, sentiment analysis & \cellcolor{darkgreen}Predicting users' mental health status and sentiment analysis& \cellcolor{darkgreen}Available  &\cellcolor{darkgreen}Centralized  \\ 
\cline{2-7}

 & \cellcolor{lightgreen} National Comorbidity Survey \cite{ncs}  & \cellcolor{lightgreen}Demographics, mental health
indicators, behavioral factors, treatments, social and environmental
factors  &\cellcolor{lightgreen} Mental health status  &  \cellcolor{lightgreen}Predicting individuals' mental health status, clustering individuals based on similar
mental characteristics   & \cellcolor{lightgreen}Public-use datasets
available/restricted-use data
available by request &\cellcolor{lightgreen}Deentralized \\ \cline{2-7}

 & \cellcolor{darkgreen}National Survey on Drug Use and Health  \cite{nsduh}  & \cellcolor{darkgreen}Demographics,  physical and mental health factors, treatments,
social and environmental factors   & \cellcolor{darkgreen}Mental health status, suicidal ideation indicator    & \cellcolor{darkgreen}Predicting individauls' mental health status, clustering individuals based on similar mental characteristics    & \cellcolor{darkgreen}Public-use datasets available/restricted-use data available by request &\cellcolor{darkgreen}Centralized \\
\cline{2-7}

       &\cellcolor{lightgreen} NLP Mental Health Conversations  \cite{nlp}  & \cellcolor{lightgreen}Conversations between users and experienced psychologists   & \cellcolor{lightgreen}Mental health status, sentiment analysis   & \cellcolor{lightgreen}Predicting individuals' mental health status and sentiment analysis, predicting personalized advice  & \cellcolor{lightgreen}Available  &\cellcolor{lightgreen}Centralized  \\
       
\cline{2-7}
 & \cellcolor{darkgreen}Nurse Stress Prediction Wearable Sensors  \cite{nspw}  &\cellcolor{darkgreen} Electrodermal activity, heart rate, skin temperature, survey responses   &\cellcolor{darkgreen} Stress level   &\cellcolor{darkgreen}  Predicting nurses' stress levels, clustering nurses based on similar mental characteristics& \cellcolor{darkgreen}Available    &\cellcolor{darkgreen} Decentralized \\ \cline{2-7}

    & \cellcolor{lightgreen} Stress in America Survey  \cite{sas}  & \cellcolor{lightgreen}Demographics, sources of stress, stress level, coping mechanisms, overall
health, behavioral factors  & \cellcolor{lightgreen}Physical and mental health status, coping effectiveness  & \cellcolor{lightgreen}Predicting individuals' stress levels,  clustering people with similar stress profiles (e.g., similar 
coping mechanisms and stress sources)  & \cellcolor{lightgreen}Available by request &\cellcolor{lightgreen} Centralized   \\ \cline{2-7}

   
 

  &\cellcolor{darkgreen} WESAD \cite{schmidt2018introducing}  & \cellcolor{darkgreen} Demographics, heart rate, skin conductance, skin temperature, blood volume pulse, respiratory rate, body temperature (sensor data)  & \cellcolor{darkgreen} Stress level  & \cellcolor{darkgreen}Predicting individuals' stress levels  & \cellcolor{darkgreen}Available  &\cellcolor{darkgreen}Decentralized  \\ 
     \hline
     \hline
 
\multirow{3}{*}{\parbox[c][4 cm][c]{0.1cm}{\centering \rotatebox{90}{\textbf{Psychiatric Data }}}}

&\cellcolor{lightred}Adolescent Brain Cognitive Development (ABCD) \cite{abcd}  & \cellcolor{lightred}Demographics, neuroimaging data, mental health indicators, academic performance, sleep patterns, cognitive test scores &\cellcolor{lightred}  Mental health status, academic performance   & \cellcolor{lightred}  Predicting individuals' mental health status or academic performance, clustering individuals based on similar mental characteristics   & \cellcolor{lightred}Available by request &\cellcolor{lightred} Decentralized  \\ \cline{2-7}

    & \cellcolor{darkred} DAIC-WOZ  \cite{daic}  & \cellcolor{darkred}Demographics, audio recordings, transcripts, facial expressions, mental health indicators  &\cellcolor{darkred} Depression level, emotion categories, engagement and distress level   &  \cellcolor{darkred}Predicting participants mental health and emotional state, clustering individuals' based on similar mental or emotional characteristics   & \cellcolor{darkred}Available by request  &\cellcolor{darkred}Centralized \\ \cline{2-7}
   & \cellcolor{lightred} UK Biobank \cite{ukb}  & \cellcolor{lightred}Demographics, imaging data, mental health indicators, environmental factors, cognitive tests, genetic data, behavioral factors, clinical records &\cellcolor{lightred}  Mental health status   & \cellcolor{lightred}  Predicting individuals' mental health status, clustering individuals based on similar mental characteristics   & \cellcolor{lightred} Available by request &\cellcolor{lightred} Decentralized  \\ \hline

\end{tabular*}

\vspace{-3mm}
\end{table*}
\vspace{-2mm}
\subsection{\underline{\textbf{Long-Term Vision 4:}} Large Language Models (LLMs)}
Research has shown that daily interactions with AI -- such as apps asking a simple ``How are you?" -- can positively influence individuals' mental well-beings \cite{bakker2018engagement} even without outcomes generated by LLMs. LLMs thus have potentials to amplify this by functioning as virtual personal counselors, delivering tailored linguistic responses and emotional support.

Although the integration of LLMs within FL has been recently studied~\cite{kuang2024federatedscope}, such integrations in mental health domain within educational settings are unexplored. Deploying LLMs in educational settings within FL offers significant benefits, such as the creation of personalized chatbots that provide tailored support to students. This approach can combine students' mental health data (e.g., StudentLife dataset) with therapy session records (e.g., NLP Mental Health Conversations and AnnoMI datasets), offering more customized assistance. Additionally, incorporating students' textual data from social media platforms (e.g., eRisk, CLPsych 2015 and Mental Health Support Feature Analysis datasets) and private institutional forums can enhance LLM performance by providing richer  contextual insights. Further, applying LLMs within FL across various clients (e.g., data from multiple schools) and comparing the results to those from LLMs applied to a single client could provide valuable insights into the advantages of FL in such environments, which is worth studying. 

However, employing LLMs in this domain faces \textit{significant ethical considerations}. Due to the sensitivity of mental health, accountability for the outcomes generated by LLMs and their impacts on individuals is unclear, raising ethical concerns. Also, it is expected that in this domain LLMs require supervision from experts (e.g., psychiatrists) during their training phase to ensure the generation of  appropriate outcomes. Thus, one crucial question in this context is determining \textit{when} and \textit{where} LLMs should be allowed to operate without human supervision.

\vspace{-2mm}
\subsection{\underline{\textbf{Long-Term Vision 5:}} Explainable AI (XAI)}

Using XAI methods within FL has been explored to enhance model interpretability \cite{barcena2022fed}; however, such usages are unexplored in the mental health domain within educational settings. Specifically, the sensitivity of  mental health  diagnoses and providing personalized treatments necessitates interpretability in ML methods. XAI can address this by making the underlying mechanisms of models transparent to the experts. Specifically, in the field of education, model transparency is crucial for decision-makers to trust the models and identify the most critical data features -- via XAI techniques such as SHapley Additive explanations (SHAP) --  in order to shape policies aimed at alleviating student stressors. Further, the knowledge of the importance of features can help universities and researchers shape mental health surveys by forming more detailed questionnaires on key features (e.g., family environment).
When integrating XAI with FL in mental health analysis is considered, exploring XAI within the VFL framework (Sec.~\ref{sec:VFL}) could unlock deeper insights into the most important features held by various entities, such as schools and clinics. Additionally, implementing XAI on decentralized datasets within a PFL framework (Sec.~\ref{sec:PFL}) could give insights into the features that play a significant role in reducing bias of various demographics during the fine-tuning processes of PFL.

It is worth mentioning that, applying XAI in the mental health domain involves some challenges. The complex and high-dimensional nature of mental health data, influenced by numerous interconnected factors, complicates the generation of clear and meaningful explanations. Implementing dimensionality reduction methods, such as Principal Component Analysis (PCA), alongside XAI can alleviate this complexity \cite{khanal2023explaining}. Additionally, privacy concerns are heightened in this context, as XAI methods may inadvertently reveal sensitive information. For example, SHAP values could disclose that low family income is a key factor in poor mental health at a specific school, potentially exposing socioeconomic details of specific students or general demographics of a school. Striking the right balance between model interpretability and student confidentiality is thus worth exploring. 




\vspace{-2mm}
\subsection{\underline{\textbf{Long-Term Vision 6:}} Multi-Modal FL (MMFL)}\label{sec:MMFL}
When implementing FL, datasets (see Tables \ref{tab:datasets} and~\ref{table:IV}) often consist of multiple data modalities. For instance, in the mental health domain, a clinic may collect textual data (e.g., patient histories), image data (e.g., medical scans), and audio data (e.g., recordings of therapy sessions). Multi-modal ML has emerged to address the challenges of training models on  diverse modalities and has been recently integrated into FL \cite{che2023multimodal}, where very few studies have taken steps toward applying MMFL in human-centered domains like healthcare \cite{borazjani2024multi}. 

In the mental health domain, datasets, such as UK Biobank and DAIC-WOZ, contain different modalities of data. MMFL thus offers a promising, yet underexplored, approach for leveraging these diverse datasets. One key challenge in this setting is that different institutions may not have access to all mental health-related data modalities -- some may only have text and images, while others may include audio. This heterogeneity of data modalities in addition to the differences in the convergence speeds of local models for each modality can disrupt the overall training process of MMFL, leading to inefficiencies in the global model's performance. Addressing these limitations is an intriguing future direction.



\vspace{-2mm}
\subsection{\underline{\textbf{Long-Term Vision 7:}} Federated Unlearning} \label{sec:unlearning}

Privacy laws such as European Union's General Data Protection Regulation (GDPR) \cite{eu-269-2014} and California Consumer Privacy Act (CCPA) in the USA \cite{ccpa} affirm users' rights to have their data erased -- known as the ``right to be forgotten." This has sparked interest in \textit{machine unlearning} \cite{bourtoule2021machine} and in turn \textit{federated unlearning} \cite{halimi2022federated}, aiming to remove the influence of specific clients' data from trained ML models without complete model retraining. 
Although few studies have explored federated unlearning for healthcare \cite{zhong2024federated}, its application in the mental health domain within educational settings is unexplored. In mental health, unique challenges arise, e.g., individuals may request the removal of only certain parts of their data while retaining others. For instance, in VFL (Sec.~\ref{sec:VFL}) a student may request the deletion of only their educational data after graduation or their clinical records once treatment is completed.
While federated unlearning makes such data removal possible, integrating it into the VFL framework to eliminate the influence of specific features is unexplored. Further, in MMFL (Sec.~\ref{sec:MMFL}), where different data modalities have distinct impacts on the model, when a student only requires removal of some modalities of data (e.g., video), the efficient implementation of federated unlearning remains to be explored. 

\vspace{-2mm}
\subsection{\underline{\textbf{Long-Term Vision 8:}} Security and Privacy} \label{sec:privacy}
In addition to enforcing the right to be forgotten -- addressed in Federated Unlearning in Sec.~\ref{sec:unlearning} -- data privacy regulations, such as GDPR and CCPA, further mandate strict protections to prevent user data exposure. While raw data remains on local devices in FL, model updates that are transferred to a central server may inadvertently reveal sensitive information, as adversaries can potentially exploit them through \textit{model inversion attacks}~\cite{geiping2020inverting} to gain insights into the local data of the clients. Consequently, to meet the privacy regulations, researchers have explored various privacy-preserving methods, including \textit{differential privacy} and \textit{cryptographic techniques}, within the FL framework. In particular, differential privacy, one of the most widely adopted techniques in FL, enhances data privacy by adding noise to the transmitted models from the clients to the server, thereby limiting the potential of precise recovering of the clients' local data~\cite{dwork2006differential}. On the other hand, cryptographic methods -- such as homomorphic and functional encryption -- enable computations (e.g., model aggregations via weighted averaging) on the encrypted transmitted models from the clients without decryption at the server, ensuring the protection of the model from the adversaries and thus confidentiality of the clients' local data~\cite{acar2018survey,boneh2011functional}. While these methods have been applied in various domains, including finance~\cite{basu2021privacy} and healthcare~\cite{adnan2022federated}, their potential applications in mental health domain warrant further investigation.

In particular, applying these techniques within different FL frameworks in the student mental health domain opens up promising research avenues. For instance, in MMFL (Sec.~\ref{sec:MMFL}), data from various modalities can have different levels of privacy sensitivity. Consider a clinic that collects both student medical scans (images) and recordings of their therapy sessions (audio). While medical scans are sensitive, students may prioritize stronger privacy protections for their therapy session recordings. Additionally, in MMFL setting, local models often use modality-specific encoders to extract features from various data modalities, such as text, images, or audio. Once extracted, these features are fused and passed through a decoder/classifier to generate the local model’s outputs. This creates an opportunity to apply differential privacy to each modality's encoder, using modality-specific noise levels tailored to the sensitivity of each data modality. However, while higher noise levels may enhance privacy, they can degrade local model performance, which in turn negatively affects the global model’s accuracy. A similar issue arises with cryptographic methods such as homomorphic encryption, where unequal encryption schemes may be required across modalities. Therefore, tuning noise levels in differential privacy or adjusting encryption in cryptographic methods for each modality are open areas of research. 

Moreover, in VFL (Sec.~\ref{sec:VFL}), where student data is split by features across different institutions, the sensitivity of the data can vary significantly. For instance, clinical data from healthcare institutions is typically more sensitive than academic records from schools. To address this, differential privacy must be customized to reflect the sensitivity of each institution's data. Protecting clinical data may require higher noise levels, but this can degrade the performance of the clinic's local model, ultimately affecting the accuracy of the global model. Finding the optimal noise levels for each institution,  to balance privacy protection with local and global model performance, remains as an open challenge in VFL.



\vspace{-2mm}
\subsection{\underline{\textbf{Long-Term Vision 9:}} Alternative FL Architectures}

In the context of students' mental health, 
real-world institutions such as schools and clinics may hesitate to fully trust external parties or a central server to aggregate sensitive data. Further, the use of a single server for model aggregations in the conventional FL architecture leads to the vulnerability of the system to a single point of failure (SPoF)~\cite{gabrielli2023survey,kang2020reliable}.
These challenges create an opportunity to explore alternative FL architectures that prioritize privacy while fostering collaboration. One particularly promising solution is \textit{decentralized FL}, which eliminates the reliance on a central server. Instead, it allows institutions to communicate and collaborate directly in a serverless environment~\cite{lalitha2018fully,qu2021decentralized,ye2022decentralized,beltran2023decentralized}. By facilitating direct interactions among institutions through peer-to-peer communications, decentralized FL enhances privacy and minimizes the risk of a single point of failure.

Another promising alternative FL architecture is \textit{semi-decentralized FL} \cite{lin2021semi,hosseinalipour2022multi,parasnis2023connectivity}. Semi-decentralized FL aims to strike a balance between (i) computation and communication efficiency and (ii) privacy protection through a hybrid model aggregation approach. In semi-decentralized FL, trusted clients are grouped into clusters, where they interact directly through peer-to-peer communications to locally aggregate their models before a few clients from each cluster engage in sending the aggregated cluster parameters to a central server~\cite{lin2021semi}. For example, upon adaptation of semi-decentralized FL to our setting of interest, regional trusted institutions -- such as schools managing academic data and clinics handling medical records -- can exchange their local models within their clusters (i.e., their trusted parties), with the central server receiving the models from only a subset of clients
to create a global model. However, determining optimal clustering strategies -- such as which clients to group, how to determine the level of trusts between the clients, and the ideal cluster size -- remains an important area for further research in the mental health domain.

Another FL architecture that is particularly effective when considering large-scale distributed model training is \textit{Hierarchical FL (HFL)}, which proposes a learning architecture consisting of  multiple tiers/layers. Upon adaptation of  HFL to our scenarios of interest, HFL will entail having institutions sending their local models to the regional edge servers, which aggregate the models and forward them to higher-tier servers (e.g., local cloud servers)~\cite{abad2020hierarchical}. This layer-wise aggregation approach enhances the scalability of the system. 
A promising and largely unexplored aspect of HFL in the mental health domain is its potential for multi-layer personalization. Achieving this multi-layer personalization would require the integration of PFL -- discussed in Sec. \ref{sec:PFL} -- with HFL, representing the first exploration of this combined approach in the mental health literature. Integrating these methods enables the creation of customized models at each tier of the network hierarchy (e.g., at the edge server level), allowing them to be tailored to meet the diverse needs of local institutions and regions.
Further, adapting differential privacy and cryptography techniques -- discussed in Sec. \ref{sec:privacy} -- to secure local models within an HFL framework has not yet been explored in the mental health domain. These privacy-preserving methods are essential for ensuring the confidentiality of sensitive student data, and their integration presents valuable opportunities for developing both personalized and secure mental health solutions through HFL.



\vspace{-2mm}
\subsection{\underline{\textbf{Long-Term Vision 10:}} FL under Data and Concept  Drift}  \label{sec:drift}
Due to the sensitive nature of mental health, inaccurate predictions can have profound consequences, emphasizing the critical need for high accuracy in FL-driven methods for early interventions and clinical treatments. To meet this need, FL must adapt to the dynamic and evolving nature of mental health data in order to be effectively applied in real-world settings. However, conventional FL frameworks often assume that the distribution of the training local data for each client is as the same as the unseen test data corresponding to this client, which is a significant limitation in practice. For example, consider a conventional FL framework used by several schools to develop a global model that predicts students' depression levels based on features such as study habits, social engagement, and academic performance. Assume that this model is trained on data collected before the COVID-19 pandemic and achieves high accuracy. However, when the pandemic arrives, significant changes occur in students' lives, leading to a shift in the data distribution (e.g., fewer social interactions and higher depression levels). Additionally, the relationship between features and depression evolves, with academic stress may become more strongly linked to depression due to the transition to online learning. As a result, the patterns observed in the pre-pandemic data may no longer hold, and the trained model may perform poorly. This situation exemplifies the impact of \textit{data drift}, where the distribution of features (e.g., study habits, social engagement) or labels (e.g., depression levels) changes over time, and \textit{concept drift}, where the relationship between features and labels evolves, on FL.

To address the challenges arising from data and concept drift, online/dynamic FL is proposed for real-time training where incoming data is continuously streamed and the model needs to detect the drift in the data and re-train to adapt~\cite{hosseinalipour2023parallel}. In the context of students' mental health data, however, some drifts in specific variables may be negligible and not require model re-training. For instance, in VFL (Sec.~\ref{sec:VFL}), clinics capture students' medical data, such as flu diagnoses. During the summer, the distribution of flu cases among students may differ significantly from that in the winter months. If the model is trained on summer data, a drift in the distribution of flu-related medical data could occur when winter arrives, resulting in a shift in the data distribution. While this represents a form of data drift, it may not necessarily require re-training the model, as flu cases may not be a major predictor of mental health status (e.g., depression levels) in this context. Therefore, this drift in flu data distribution does not substantially affect the model’s ability to predict mental health outcomes and could be safely overlooked. On the other hand, even a minor shift in the distribution of academic factors, such as an increase in the number of exams during the semester, could significantly impact students' mental health (e.g., leading to higher stress levels). Since academic stress is a major factor influencing mental health, this type of data drift should definitely be captured, and the model should be re-trained to adapt to these new patterns. Developing mental health tailored drift detection algorithms within online/dynamic FL frameworks are thus necessary to identify specific data drifts, assess their impact on model performance, and decide on model re-training, which remains to be an open area of research.








\vspace{-2mm}
\section{Conclusion}

\noindent In this paper, we explored the transformative potential of FL in advancing mental health analysis within the education domain while preserving data privacy and alleviating ethical concerns. By reviewing key mental health issues affecting students, including anxiety, stress, depression, ADHD, and SUD, we highlighted how these conditions impair academic performance and overall well-being. We then examined current applications of ML techniques in mental health analysis and identified the limitations posed by centralized ML frameworks, particularly regarding privacy risks and data aggregation challenges.
To address these concerns, we studied FL as a promising solution that leverages decentralized data from educational institutions, mental healthcare providers, and other relevant sources. FL’s capacity to perform privacy-preserving model training without transferring raw data presents a paradigm shift in mental health research, enabling collaborative analytics while ensuring data confidentiality. By summarizing current FL applications in mental health, we highlighted both promising advances and existing gaps, particularly in the educational context where FL adoption remains limited.
Additionally, we outlined a research roadmap that categorizes future directions into short- and long-term goals. In the short-term, we proposed extending FL methods to heterogeneous and decentralized datasets while addressing data heterogeneity, communication efficiency, and fairness. In the long-term, we envisioned innovations such as VFL for integrating data from multiple institutions, complementary server-side learning, personalized and multi-task FL, and XAI. We further emphasized emerging technologies such as LLMs for personalized mental health counseling and federated unlearning for adhering to regulatory frameworks like GDPR.
Moreover, we discussed security and privacy enhancements using advanced cryptographic techniques and differential privacy methods, which are critical for protecting sensitive mental health data. We also proposed alternative FL architectures, such as decentralized and semi-decentralized models, and considered adapting FL to dynamic environments characterized by data and concept drift.
 Our proposed directions aim to push the boundaries of privacy-conscious AI/ML applications, ensuring both data security and actionable insights.

\bibliographystyle{ieeetr}
\bibliography{example_paper}

\begin{IEEEbiography}
[{\includegraphics[width=1in,height=1.25in,clip,keepaspectratio]{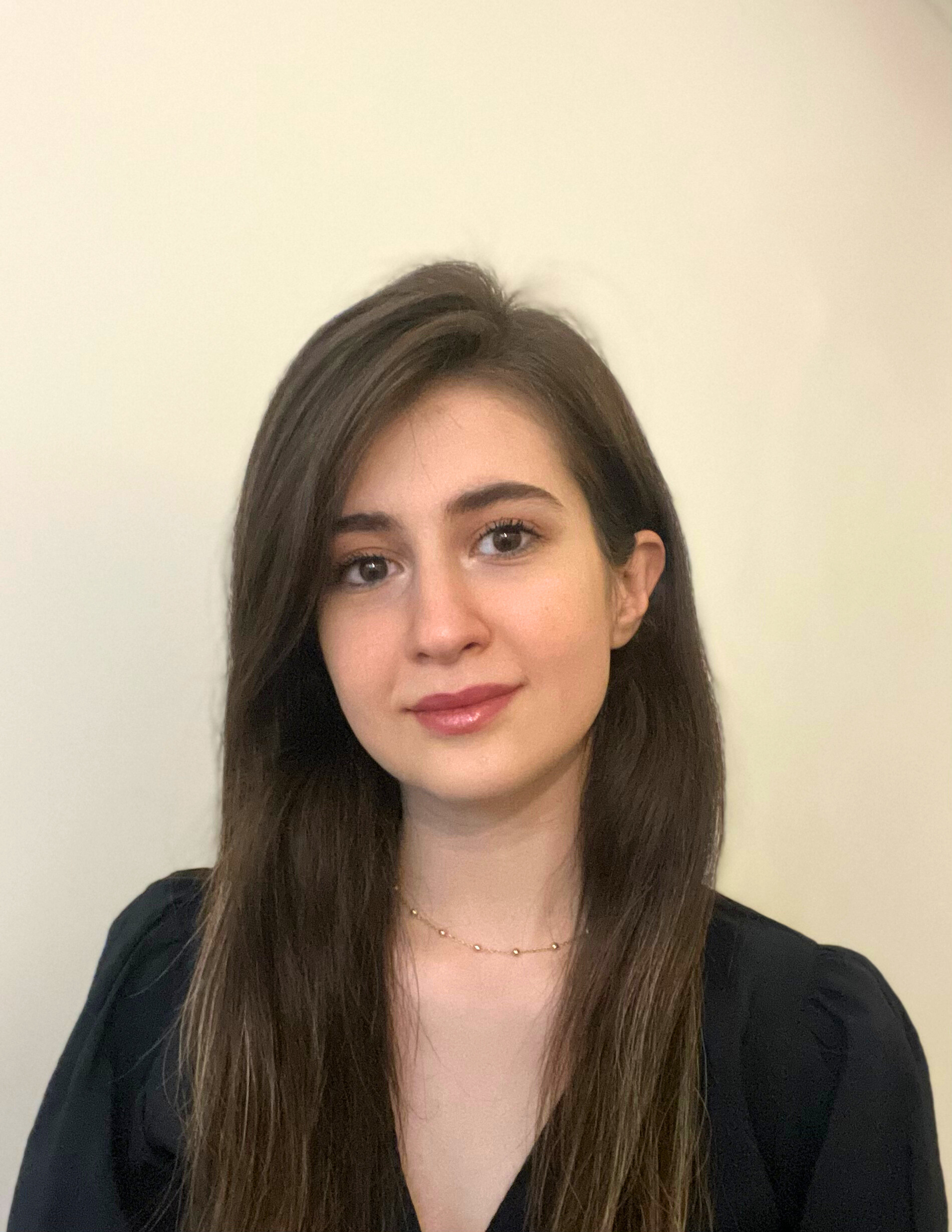}}]{Maryam Ebrahimi} received the B.S. degree in applied mathematics from Sharif University of Technology, Tehran, Iran in 2024. She has been the recipient of \textit{Bronze Medal of National Iranian Olympiad in Mathematics}, 2018. Her research interests include machine learning in education, distributed machine learning, privacy, and game theory.

\end{IEEEbiography}

\begin{IEEEbiography}
[{\includegraphics[width=1in,height=1.25in,clip,keepaspectratio]{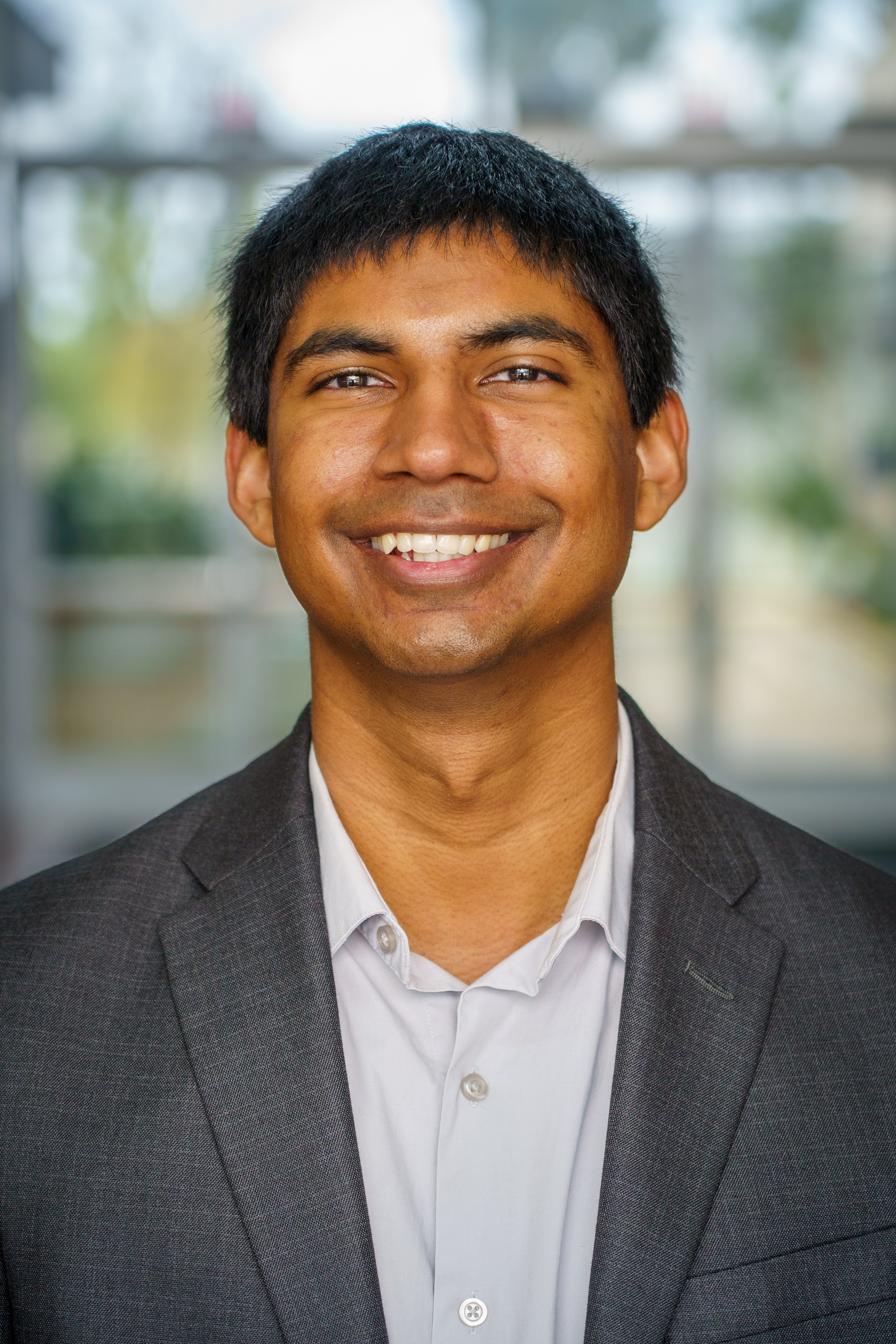}}]{Rajeev Sahay} received the B.S. degree in electrical engineering from The University of Utah, Salt Lake City, UT, USA, in 2018, and the M.S. and Ph.D. degrees in electrical and computer engineering from Purdue University, West Lafayette, IN, USA, in 2021 and 2022, respectively. Currently, he is a faculty member in the Department of Electrical and Computer Engineering at UC San Deigo. He was the recipient of the Purdue Engineering Dean’s Teaching Fellowship and was named an Exemplary Reviewer by the IEEE Wireless Communications Letters. His research interests lie in the intersection of networking and machine learning, especially in their applications to wireless communications and engineering education. 
\end{IEEEbiography}
\vspace{-3mm}

\begin{IEEEbiography}
[{\includegraphics[width=1in,height=1.25in,clip,keepaspectratio]{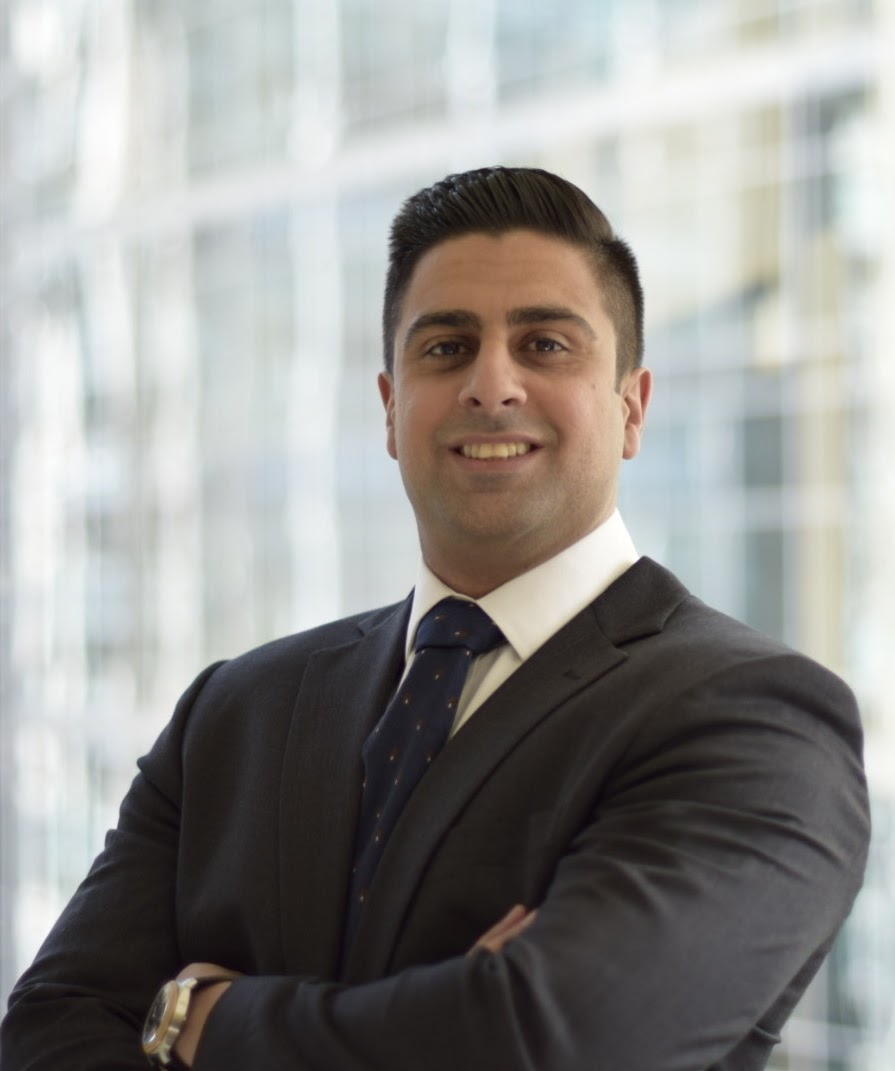}}]{Seyyedali Hosseinalipour}[M] received the B.S. degree in electrical engineering from Amirkabir University of Technology, Tehran, Iran, in 2015 with high honor and top-rank recognition. He then received the M.S. and Ph.D. degrees in electrical engineering from North Carolina State University, NC, USA, in 2017 and 2020, respectively; and was a postdoctoral researcher at Purdue University, IN, USA from 2020 to 2022. 
He is currently an assistant professor in the department of electrical engineering, University at Buffalo-SUNY, NY, USA. 
He was the recipient of the \textit{ECE Doctoral Scholar of the Year Award} (2020) and \textit{ECE Distinguished Dissertation Award} (2021) at NC State University; and \textit{Students’ Choice Teaching Excellence Award} (2023) at University at Buffalo--SUNY. Furthermore, he was the first author of a paper published in IEEE/ACM Transactions on Networking that received the \textit{2024 IEEE Communications Society William Bennett Prize}.
He has served as the TPC Co-Chair of workshops/symposiums related to machine learning and edge computing for IEEE INFOCOM, GLOBECOM, ICC, CVPR, ICDCS, SPAWC, WiOpt, and VTC. He has also served as the guest editor of IEEE Internet of Things Magazine for the special issue on \textit{Federated Learning for Industrial Internet of Things} (2023).
His research interests include the analysis of
modern wireless networks, synergies between machine learning methods and fog/edge
computing systems, distributed/federated machine learning, and network optimization.
\end{IEEEbiography}
\vspace{-3mm}
\begin{IEEEbiography}
[{\includegraphics[width=1in,height=1.25in,clip,keepaspectratio]{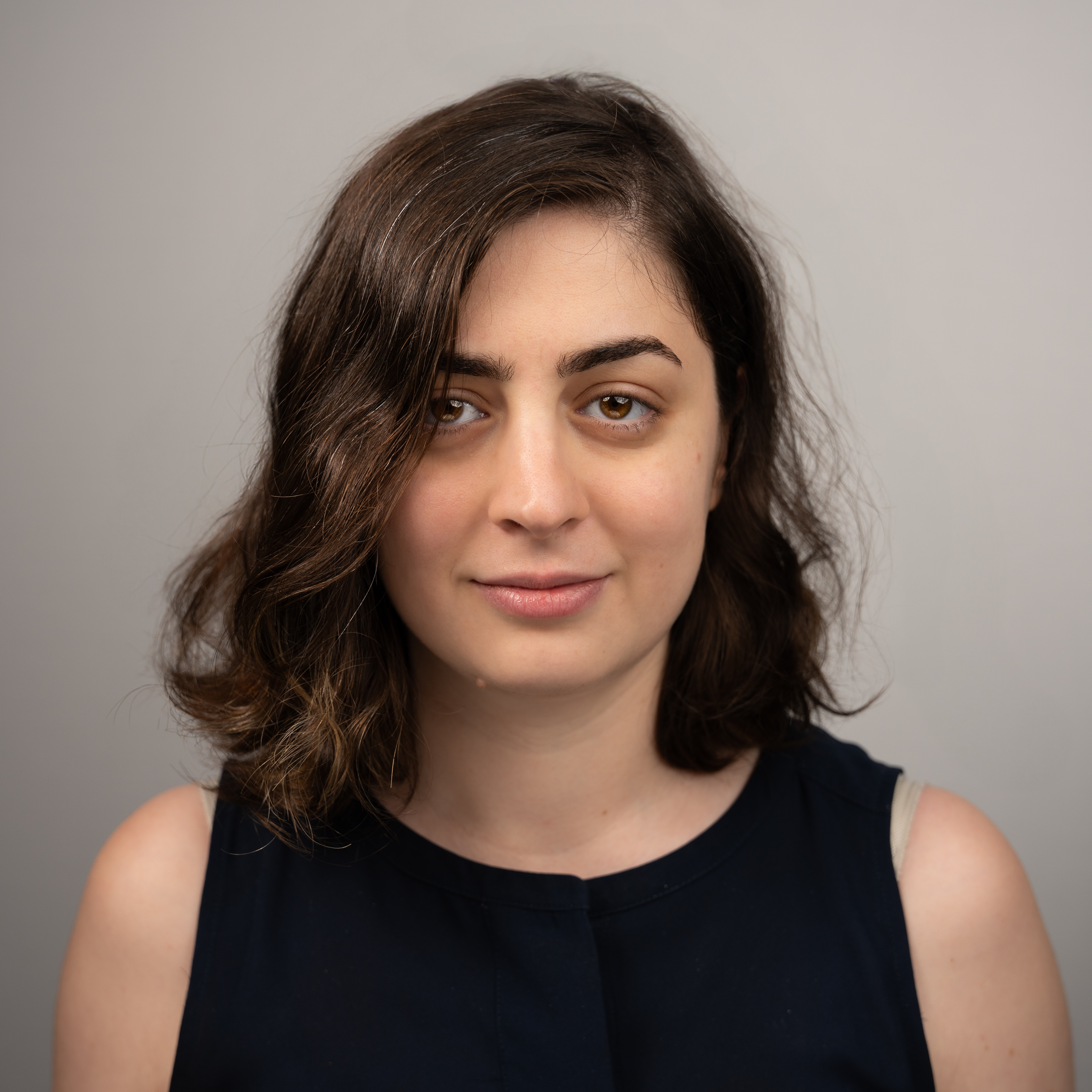}}]{Bita Akram} received her B.S. degree in Computer Engineering from Sharif University of Technology, Tehran, Iran, in 2013. She then earned her M.Sc. in Computer Science from the University of Calgary, Calgary, Canada, in 2015, and her Ph.D. degree from North Carolina State University, Raleigh, USA, in 2019. Dr. Akram is currently an Assistant Professor in the Department of Computer Science at North Carolina State University. Her research lies at the intersection of Artificial Intelligence, Learning Sciences, and Human-Computer Interaction, where she explores innovative approaches to enhance education through AI and technology. Her work is supported by several NSF-funded projects. She has been recognized with multiple SIGCSE Best Paper Awards and actively contributes to the research community. Dr. Akram has served on program committees for leading conferences in educational data mining and computer science education, including EDM, EAAI, SIGCSE, and ITICSE. Additionally, she has held roles such as Workshop Co-Chair for EDM and CSEDM.

\end{IEEEbiography}

\end{document}